\documentclass[11pt]{IEEEtran}

\usepackage[margin=1in]{geometry}

\usepackage{amsmath,amssymb}
\usepackage[fleqn]{mathtools}
\usepackage{thmtools}

\usepackage{mathtools}
\mathtoolsset{centercolon}

\usepackage{multicol}

\usepackage[colorlinks = true]{hyperref}
\usepackage{xcolor}
\definecolor{darkred}  {rgb}{0.5,0,0}
\definecolor{darkblue} {rgb}{0,0,0.5}
\definecolor{darkgreen}{rgb}{0,0.5,0}
\hypersetup{
  urlcolor   = blue,         
  linkcolor  = darkblue,     
  citecolor  = darkgreen,    
  filecolor  = darkred       
}

\usepackage{cleveref}
\crefname{lemma}{Lemma}{Lemmas}
\crefname{proposition}{Proposition}{Propositions}
\crefname{definition}{Definition}{Definitions}
\crefname{theorem}{Theorem}{Theorems}
\crefname{conjecture}{Conjecture}{Conjectures}
\crefname{corollary}{Corollary}{Corollaries}
\crefname{section}{Section}{Sections}
\crefname{appendix}{Appendix}{Appendices}
\crefname{figure}{Fig.}{Figs.}
\crefname{equation}{Eq.}{Eqs.}
\crefname{table}{Table}{Tables}
\crefname{claim}{Claim}{Claims}
\crefname{fact}{Fact}{Facts}
\crefname{example}{Example}{Examples}

\newtheorem{theorem}{Theorem}
\newtheorem{proposition}[theorem]{Proposition}
\newtheorem{definition}[theorem]{Definition}

\newtheorem{lemma}[theorem]{Lemma}
\newtheorem{conjecture}[theorem]{Conjecture}

\newtheorem{fact}[theorem]{Fact}

\newcommand{\C}{\mathbb{C}}
\newcommand{\F}{\mathbb{F}}
\newcommand{\R}{\mathbb{R}}
\newcommand{\N}{\mathbb{N}}

\newcommand{\M}{\mathcal{M}}
\newcommand{\X}{\mathcal{X}}
\newcommand{\E}{\mathcal{E}}
\renewcommand{\H}{\mathcal{H}}
\newcommand{\qLDPC}{{\sf{qLDPC}}}

\newcommand{\be}{\begin{eqnarray}}
\newcommand{\ee}{\end{eqnarray}}

\DeclarePairedDelimiter{\set}{\lbrace}{\rbrace}

\DeclarePairedDelimiter{\floor}{\lfloor}{\rfloor}

\newcommand{\ket}[1]{|#1\rangle}
\newcommand{\bra}[1]{\langle#1|}

\newcommand{\mnote}[1]{}

\newcommand{\eps}{\varepsilon}
\renewcommand{\epsilon}{\varepsilon}

\newcommand{\x}{\otimes}
\newcommand{\xp}[1]{^{\otimes #1}}

\newcommand{\ct}{^{\dagger}}
\newcommand{\tp}{^{\mathsf{T}}}

\DeclareMathOperator{\im}{im}     
\DeclareMathOperator{\syst}{syst} 

\begin{document}
\onecolumn
\title{The Need for Structure in Quantum LDPC Codes}

\author{Lior Eldar
\thanks{Lior Eldar is with the Center for Theoretical Physics, MIT} \and
Maris Ozols
\thanks{Maris Ozols is with Department of Applied Mathematics and Theoretical Physics, University of Cambridge} \and
Kevin Thompson
\thanks{Kevin Thompson is with School of Engineering and Applied Science, Harvard}}
\maketitle

\begin{abstract}
Existence of quantum low-density parity-check (LDPC) codes whose minimal distance scales linearly with the number of qubits 
is a major open problem in quantum information. 
Its practical interest stems from the need to protect information in a future quantum computer,
and its theoretical appeal relates to a deep ``global-to-local'' notion in quantum mechanics: 
whether we can constrain long-range entanglement using local checks.  Given the inability of lattice-based quantum LDPC codes to achieve linear distance \cite{BT09}, 
research has recently shifted to the other extreme end of topologies, 
so called high-dimensional expanders \cite{EK16}.  In this work we show that trying to leverage the mere ``random-like'' property of these
expanders to find good quantum codes may be futile: 
quantum CSS codes of $n$ quits built from
$d$-complexes that are $\eps$-far from perfectly random, 
in a well-known sense called discrepancy, have a small minimal distance.  Quantum codes aside, our work places a first upper-bound on the systole of high-dimensional expanders with small discrepancy,
and a lower-bound on the discrepancy of skeletons of Ramanujan complexes due to Lubotzky~\cite{Lubotzky2014}.

\end{abstract}



\section{Introduction}

\subsection{General}

\emph{Quantum error-correcting codes} (QECCs) and \emph{high-dimensional expanders}
are two distinct fields, with somewhat disparate motivations.
Quantum error correction studies the ability to efficiently encode quantum information in a way that is robust against errors induced by environment \cite{QEC,NC11}.
The theoretical study of quantum error correction is central to our understanding of multi-particle entanglement,
and in more practical terms, existence of quantum codes with sparse parity checks, i.e. quantum LDPC codes $(\qLDPC)$ would likely facilitate the construction of quantum computers \cite{Got13}.
On the other hand, the nascent study of high-dimensional expanders (see for example \cite{Lubotzky2014}) is motivated by an attempt to replicate the enormous
success of $1$-dimensional expanders, namely \emph{expander graphs}, in theoretical computer science and mathematics more broadly \cite{HLW06,Lub12}.
In this field, one attempts to provide a ``natural'' generalization of expansion to hyper-graphs in a way that inherits the fundamental properties of expander graphs,
perhaps the most prominent of these
being the tight connection between spectral properties (spectral gap) and combinatorial properties (Cheeger constant) \cite{Cheeger, A86}

When considering high-dimensional expanders we
refer to the notion of a \emph{chain complex} from homological algebra. 
For a given dimension $d$, a chain complex $\X = (X_0, X_1, \dotsc, X_d)$, or simply a \emph{$d$-complex}, is a sequence of linear spaces $X_i$ (over $\F_2$) together with linear maps $\partial_i : X_i \to X_{i-1}$ between them, known as
\emph{boundary operators} satisfying $\partial_{i-1} \circ \partial_i = 0$ for each $i$.
It is well-known for a while now \cite{K97,BK01} that a $2$-complex gives rise - in a fairly natural way - to a certain QECC called CSS code (see \cref{sec:CSSx} for more details).

Recent studies by Kaufman, Kazhdan and Lubotzky \cite{KKL}, and subsequent work by Evra and Kaufman \cite{EK16} have established
that certain efficiently constructible $d$-dimensional complexes, called Ramanujan complexes, 
have $d-1$ skeletons (the set of all $d-1$ faces contained in some face of the $d$ complex) that
possess a property called
co-systole expansion.
This property generalizes the notion of combinatorial expansion in expander graphs and, perhaps more generally, the notion
of local testability of combinatorial properties \cite{KL14}.

On the other hand, recent studies in quantum information \cite{FH13, BH13, EH15, Has16} have revealed that 
the topology underlying a locally-defined quantum code plays a crucial role in the behavior of its
long-range entanglement.
In more concrete terms, these works suggest that deviating from complexes defined on regular grids towards 
more random-looking topologies may give rise to locally-defined quantum systems whose ground-space entanglement
is robust even against {\it adversarial} noise.



A central conjecture in this context is the quantum LDPC conjecture ($\qLDPC$)
which posits the existence of quantum LDPC codes whose minimal distance scales linearly in the number
of qubits (see \cref{qLDPC}).
In terms of topology this requires the existence of a $2$-complex of bounded degree with a systole {\it and} co-systole
both scaling linearly in the number of $1$-faces.

Here too, since 
complexes with large systole / co-systole 
cannot arise from a cellulation of a low-dimensional manifold \cite{BT09}, it is then
natural to conjecture that high-dimensional expanders may give rise to such behavior \cite{EK16,Has16}.
Yet, in this work, we provide some negative evidence to this intuition.
We show that trying to leverage the mere random-looking property of high-dimensional expanders to 
construct quantum LDPC codes with large distance may be futile.

To formally connect quantum codes and complex chains
we refer to the standard map from $2$-complexes to quantum error correcting codes known as homological CSS codes \cite{BD07}:
We consider a $2$-complex $\X = (X_0,X_1,X_2)$ with boundary maps $\partial_2: X_2 \to X_1$ and $\partial_1: X_1 \to X_0$.
We will associate the stabilizers of the quantum code to the images of the maps $\partial_2$, $\partial_1^T$.  For instance we will assign the rows of $\partial_1$ to (a generating set of) Pauli-$Z$ check operators,
and the columns of $\partial_2$ to (a generating set of) Pauli-$X$ check operators.  This can be done by mapping each binary vector
$b \in \im(\partial_1^T) \in \F_2^n$ to $\bigotimes_{i=1}^n Z^{b(i)}$ (and the corresponding map for $X$ operators).

We then consider families of such $2$-complexes of \emph{bounded degree} $K_1, K_2 = O(1)$, i.e.\ the Hamming weight of each column of $\partial_2$ and each row of $\partial_1$
is $K_2$ and $K_1$ respectively.
We will use the notation $d = O(1)$ to denote the maximum Hamming weight of each row of $\partial_2$ and column of $\partial_1$, i.e. the \emph{locality}
of the Pauli stabilizers defining the CSS code.
Our main theorem relates a well-known notion of pseudo-randomness (see \cref{def:ePR}) and shows that the natural CSS quantum code
arising from an $\eps$-pseudo-random $2$-complex has a stringent \emph{upper bound} on its minimal distance:

\begin{theorem}\label{thm:main}
Let $\{\X_n\}_{n\in \N}$ be a family of $2$-complexes
\begin{equation}\label{eq:Xn}
  \X_n : X_2 \overset{\partial_2}{\longrightarrow} X_1 \overset{\partial_1}{\longrightarrow} X_0,
\end{equation}
where each column of $\partial_2$ and each row of $\partial_1$ has weight $d = O(1)$,  each row of $\partial_2$ has
weight $K_2 = O(1)$, and each column of $\partial_1$ has weight $K_1 = O(1)$.
Denote by ${\cal C}(\X_n) = [[n,k,d_{\min}]]$ the quantum code associated to $\X_n$.
Let 
$$
\eps_0 = \min\left\{\left(H^{-1}(2^{- 2\log(2d)/d})\right)^{2d},2^{-2d}\right\},
$$
where $H(x)$ is the binary entropy function.
If $\X_n$ is $\eps$-pseudorandom, 
for some constant $0 < \eps \leq \eps_0$ independent of $n$,
then 
$$
d_{\min} \leq\frac{12}{d^2}\epsilon^{1/(2d)} \log^2(1/\epsilon)n.
$$
\end{theorem}

%

\subsection{Quantum error correction}\label{sec:qecc}

The study of quantum error correcting codes is driven by the high-level goal of making quantum information robust against environment-induced errors. Generally speaking, a quantum error correcting code (QECC) for encoding $k$ qubits into $n$, with $n \geq k$, is a an isometry
$$V: \H := (\C^2)\xp{k} \hookrightarrow \H' := (\C^2)\xp{n}$$
that protects quantum states by ``spreading'' them out over a larger subspace.
Specifically, for any Pauli error $\E := \E_1 \x \dotsb \x \E_n$, where each $\E_i \in \set{I,X,Y,Z}$ is one of the four Pauli matrices, in which at most $d_{\min}$ terms $\E_i$ are different from the identity matrix $I$, any two orthogonal states $\ket{\psi_1}, \ket{\psi_2} \in \H$ remain orthogonal under $V\ct \E V$: 
$$
\bra{\psi_1} V\ct \E V \ket{\psi_2} = 0.
$$
Hence a \emph{logical state} $\ket{\psi}$, after being encoded as $V \ket{\psi}$, can be recovered from any error that acts on at most $\floor{\frac{d_{\min}-1}{2}}$ qubits.
As in the study of classical error correction, one then usually looks for efficient ways to encode and decode quantum states, and to achieve optimal
parameters in terms of rate and minimal distance.

In QECCs, as in classical ECCs, one can define a code ${\cal C}$, namely a subspace of $\H'$, 
as the space stabilized by
a set of parity checks $P_i$ - namely:
a quantum state $\ket{\psi} \in {\cal C}$ if $P_i \ket{\psi}=0$ for each $i$. In classical ECCs, when the $P_i$'s are sparse, then the resulting code is said to be a low-density parity-check code, or LDPC for short.

In classical ECC, it is well known that LDPC codes with linear minimal distance can be constructed \cite{SS96},
and those can even achieve the error-correction capacity, i.e. there is (almost) no penalty for the
extra requirement of locality.
Yet in the quantum setting we do not even know how to construct an LDPC code with linear distance for any
{\it non-zero} rate.

\begin{conjecture}[$\qLDPC$ Conjecture]\label{qLDPC}
There exists a quantum code ${\cal C}$ of positive rate with a set of parity checks $P_i$, each of which can be written as 
$P_i = I_a \otimes p_i \otimes I_b$, where $p_i$ is a matrix of dimension at
most $2^K$ for $K = O(1)$, such that $d_{\min}({\cal C}) = \Omega(n)$.
\end{conjecture}

The $\qLDPC$ conjecture is not a mere question about optimality of parameters.
In a sense, it goes to the core of our understanding of multi-particle entanglement since the states corresponding to code-words of a quantum code with linear distance can be shown to be highly-entangled in a well-defined sense:
hence a $\qLDPC$ code implies that we can generate a quantum system with {\it global} entanglement
using only {\it local} constraints.

\subsection{High-Dimensional Expanders}\label{sec:Highdim}

Expander graphs are graphs with a sparse adjacency matrix that are ``rapidly-mixing'' in a well-defined sense \cite{A86, H06}.
There is no de facto standard definition for high-dimensional expanders and in the last few years, the nascent research
in this field has explored various definitions that are not all known to be comparable.
We survey here briefly the definitions that are most relevant to us and mention that more are available in the literature \cite{LM06, KKL, KL14, Par13, PRT16, EK16, Lubotzky2014}:

We begin with a definition of chain complexes, boundary maps, and homology (in our case, all linear maps are defined over $\F_2$):

\begin{definition}
A \emph{$d$-dimensional chain complex} $\X$, or \emph{$d$-complex}, is a tuple $(X_0,\dotsc, X_d)$ of $d+1$ vector spaces over $\F_2$, with linear maps $\partial_i: X_i \to X_{i-1}$ for all
$1 \leq i \leq d$, known as {boundary maps}, that satisfy the following boundary property:
\begin{equation}
  \partial_i \circ \partial_{i+1} = 0, \quad \forall i: 1 \leq i < d.
  \label{eq:bp}
\end{equation}
The space $X_0$ represents formal sums over the vertices of the complex, $X_1$ represents formal sums over the edges, etc.
We define $Z_i := \ker(\partial_i)$, known as $i$-cycles, and $B_i := \im(\partial_{i+1})$, known as $i$-boundaries. According to \cref{eq:bp}, $B_i \subseteq Z_i$ for all $i$, so we can define the $i$-th {homology} as the quotient group (space) $Z_i / B_i$.
We can also define {co-boundary maps} as $\partial^i := \partial_i\tp$.
In this case,
\begin{equation}
  \partial^{i+1} \circ \partial^i = 0, \quad \forall i: 1 \leq i < d.
\end{equation}
We define {co-cycles} and {co-boundaries} as $Z^i := \ker(\partial^{i+1})$ and $B^i := \im(\partial^i)$, respectively, and hence $B^i \subseteq Z^i$ for all $i$. The $i$-th {co-homology} is then the quotient $Z^i / B^i$.
\end{definition}

Our working definition for expansion is a natural generalization of Cheeger's constant to hyper-graphs:
\begin{definition} [$\eps$-pseudo-random complex] \label{def:ePR}
Let $\X = (X_0,\hdots, X_d)$ be a $d$-dimensional complex on $|X_0| = n$ vertices in which each ($d-1$)-cell is contained in exactly $K$ $d$-faces ($K$ regular, $d$ uniform).  
Denote the set of $d$-faces as $E$.  For any $S \subseteq [n]$ a subset of the vertices, and for any $j \leq d$, define $A(S, j)$ to be the number of $d$-faces in $E$ with $j$ members in $S$.  Then $\X$ is said to be \emph{$\eps$-pseudo-random} ($\eps$-PR) if for all subsets of the vertices, $S \subseteq [n]$ we have:
\be 
\left|\frac{A(S, j)}{|E|}-\binom{d}{j}\left( \frac{|S|}{n}\right)^j \left(1-\frac{|S|}{n} \right)^{d-j} \right|< \epsilon
\ee

\end{definition}

Hence, a complex is $\eps$-PR if the uniform distribution on its faces very nearly matches the uniform distribution over all faces of weight $d$, up to an additive error $\eps$.  The reader may note that the second term on the LHS is not really the probability that a uniformly chosen face of weight $d$ has $j$ vertices in $S$, but it is asymptotically close to this probability for large $n$ and $d=O(1)$.  
Often, we will interpret the rows of $\partial_d$ as the hyperedges of a 
$d$-uniform hyper-graph
and then say that $\partial_d$ is $\eps$-PR whenever $\X$ is $\eps$-PR.    

Our definition generalizes the notion of discrepancy defined for graphs, i.e. $2$-complexes.
For graphs, it is known that the discrepancy of the graph is characterized by the spectrum of the graph,
via the expander mixing lemma \cite{BL06}.
In the context of high-dimensional expanders our definition is very similar to the one defined by Parzanchevski in \cite{PRT16}, where a hyper-graph was said to be $\eps$-PR
if for any partition of $V$ into $d+1$ parts the fraction of hyper-edges with one vertex in each part is the same
as that fraction for a uniformly random $d$-hyper-edge, up to additive error $\eps$.
Under this definition, Parzanchevski then established a high-dimensional generalization of the expander mixing lemma, showing that simplicial complexes
with a large spectral gap (referring here to the Hodge Laplacian of simplicial complexes) are also $\eps$-PR.

\noindent
The second definition is due to \cite{KKL,EK16}, and considers the $\F_2$ expansion of non-cycles:

\begin{definition}[Systole / co-systole]
Let $\X$ be a $d$-complex with $\F_2$ boundary maps $\partial_j: X_j \to X_{j-1}$ for all $0 < j \leq d$.
Let $Z_j := \ker(\partial_j)$ and $B_j := \im(\partial_{j+1})$ denote the $j$-cycles and $j$-boundaries.
The \emph{$j$-th systole} is then defined as:
$$
\syst_j(\X) := \min_{w \in Z_j - B_j} |w|.
$$
Likewise, the \emph{$j$-th co-systole} is defined as:
$\min_{w \in Z^j - B^j} |w|$.
\end{definition}

\noindent
For cellulations of geometric manifolds, $\syst_1(\X)$ can intuitively be understood as the length of a shortest non-trivial cycle.

\begin{definition}[Cycle / co-cycle expansion]
Let $X$ be a $d$-complex with $\F_2$ boundary maps $\partial_j: X_j \to X_{j-1}$ for all $0 < j \leq d$.
$\X$ has $\eps$ \emph{$j$-th cycle expansion} if 
$$
\min_{x\in X_j - Z_j} \frac{|\partial_j x|}{\min_{w\in Z_j} |x + w|} \geq \eps.
$$
\end{definition}

In \cite{EK16} it is then shown that $d-1$ skeletons of Ramanujan $d$-complexes have simultaneously $\eps$ co-cycle expansion,
and a large (linear-size) co-systole.
Hence these complexes are said to be co-systole expanders. 
In this context, one may interpret our main theorem as a statement about high-dimensional expanders which does not use the
language of quantum codes (see \cref{sec:qecc} for relevant definitions).
For example, using the definition of systole / co-systole expansion due to Evra and Kaufman \cite{EK16}
we claim:

\begin{theorem}
[Pseudo-random complexes have bounded systoles]
Let $\{\X_n\}_{n\in \N}$ be a family of $2$-complexes, with $\X_n$ as in \cref{eq:Xn} and $\dim(X_1) = n$, 
where each column of $\partial_2$ and each row of $\partial_1$ has weight $d = O(1)$ and each row of $\partial_2$ has weight $K_2 = O(1)$
and each column of $\partial_1$ has weight $K_1 = O(1)$.
Let 
$$
\eps_0 = \min\left\{\left(H^{-1}(2^{- 2\log(2d)/d})\right)^{2d},2^{-2d}\right\},
$$
where $H(x)$ is the binary entropy function.
If $\X_n$ is $\eps$-pseudorandom, 
for some constant $0 < \eps \leq \eps_0$ independent of $n$,
then 
$$
\syst_1(\X_n) \leq \frac{12}{d^2}\epsilon^{1/(2d)} \log^2(1/\epsilon) n.
$$
\end{theorem}
Since by \cite{EK16} the $d-1$ skeletons $\X_n$ of Ramanujan $d$-complexes are co-systole expanders, then in particular the co-systole $\syst^1(\X_n) \geq c_0 n$ for some constant $c_0$.
Hence there exists $\eps' = \eps'(c_0)$ such that these skeletal complexes have discrepancy at least $\eps'$.

\subsection{CSS codes: connecting quantum codes to high-dimensional expanders}\label{sec:CSSx}

CSS codes invented by Calderbank, Shor, and Steane (see \cite{NC11} or the original papers \cite{CS96,Steane96a,Steane96b}) are one of the earliest and most influential type of quantum codes.
Arguably, their greatest advantage, is that they can be defined using pairs of {\it classical} codes, which then allows one to think of these quantum codes as classical codes with certain restrictions:

\begin{definition}[CSS code]\label{def:CSS}
A quantum $[[n,k,d_{\min}]]$ \emph{CSS code} is a pair of classical codes $C_X,C_Z \subseteq \F_2^n$ such that $C_Z \subseteq C_X^{\perp}$ and $C_X \subseteq C_Z^{\perp}$.
The parameters of the code are given by $k := n-\dim(C_X)-\dim(C_Z)$ and
$$
d_{\min} := \min_{w\in C_X^{\perp} - C_Z, C_Z^{\perp} - C_X} |w|.
$$
We denote by $m_X = \dim(C_X), m_Z = \dim(C_Z)$.
\end{definition}

The codes $C_X$ and $C_Z$ describe the $X$ and $Z$ stabilizers of the codeword states, respectively.  In the language of stabilizer quantum error correcting codes \cite{NC11}, we can specify a complete independent set of generators for the stabilizers by specifying a complete basis for the code spaces $C_X$ and $C_Z$.  

Hence, to find a good quantum CSS code, we are required to find a pair of \emph{classical} codes that contain the dual of each other, have a large quotient group (code rate), and a large value of $d_{\min}$ defined above.
In addition, if we require LDPC, we need also for each code $C_X$ and $C_Z$ to be sparsely generated.  While finding good classical LDPC codes is today an easy task, the extra condition of pairwise duality has prevented thus far construction of quantum LDPC CSS codes with minimal distance scaling linearly in the number of qubits.  

One of the many elegant features of CSS codes, that has been recognized a long time ago, 
is that CSS codes naturally arise from chain complexes that are induced by a cellulation of a manifold,
or more generally, from complex chains:
\begin{definition}[Map from $2$-complexes to CSS codes]\label{def:CSSx}
Let $\X = (X_0,X_1,X_2)$ be a $2$-complex, with boundary maps $\partial_2,\partial_1$.
The CSS code ${\cal C}$ corresponding to $X$ is defined by choosing $C_X := \im \partial_1$, $C_Z := \im \partial_2\tp$,
and fixing  ${\cal C}(\X) = (C_X, C_Z)$.
\end{definition}
Indeed, many well known constructions can be thought of in this way \cite{BD07}\cite{K97}.

\subsection{Discussion and previous work}

In this work we 
further study the connection between the topology underlying a locally-defined quantum
system and the property of the entanglement that arises from it: we
show that high-dimensional complexes that are $\eps$-pseudo-random
have actually very poor quantum code distance.
This implies that attempting to construct CSS codes in this way will probably not yield a construction of quantum 
$\qLDPC$ codes with linear distance.
Furthermore, it implies that $d$-complexes that are $\eps$-pseudo-random for small $\eps$ cannot be very
good systole / co-systole expanders.

Our result is arguably surprising in the context of high-dimensional expanders:
Evra and Kaufman recently showed \cite{EK16} that the $d-1$ skeletons of Ramanujan $d$-complexes are good co-systole expanders.
Together with our theorem this
implies a \emph{lower bound} on the discrepancy $\eps$ of these skeleton complexes.
Stepping back, our result is thus an example where the quantum perspective sheds light on questions in other fields.


In the context of quantum codes, there are no-go results for quantum codes that can be embedded into lattices.  In particular Bravyi et al. \cite{BT09, BPT10} have shown that stabilizer codes that can be embedded into $D$-dimensional lattices where $D=O(1)$, and each stabilizer is supported on $O(1)$ qubits within some hyper-cube, cannot have linear distance.  Our result can be viewed as the ``other end'' of this limit.  We show that codes which are in a sense ``strongly'' not embeddable into a lattice also have small distance.
We note, however, that our bounds can only show a small constant linear distance bound, and not a sub-linear distance bound, which would have ruled out linear distance quantum codes on very low discrepancy expanders altogether.

Interestingly, under a certain conjecture in high-dimensional geometry, Hastings \cite{Has16} recently showed a quantum CSS code with distance $n^{1-\eps}$, for arbitrarily small $\eps>0$,
whose parity check matrices have sparsity $\log(n)$,
using a cellulation of a family of random lattices, called LDA.
Given the inherent embedding in a lattice, we believe that the $2$-complex of his code would actually be very far from pseudo-random.
This suggests that perhaps the ``right'' way to resolve the $\qLDPC$ conjecture is to look for high-dimensional manifolds that avoid pseudo-randomness.

Perhaps more fundamentally, our work suggests that quantum multi-particle entanglement may be inherently limited on random-looking topologies,
contrasting the intuition stemming from the classical theory of computer science that random-looking topologies are
more ``robust''.
The notion that highly-expanding topologies are adversarial to large-scale quantum entanglement
resonates with a sequence of results
of a somewhat different context:
In \cite{BH13} the authors show that
the ground-states of
$2$-local Hamiltonians whose graph is \emph{expanding}
can be approximated by tensor-product states.
A similar result in \cite{AE15} shows this for $k$-local commuting Hamiltonians with a bipartite
form of expansion, using a more stringent criterion called {\it local expansion}.
These results impose a much more stringent structure on the quantum system
(namely $2$-locality, or local expansion) and derive not only a minimal distance upper
bound for a corresponding quantum code, but in fact an upper bound on the locality of entanglement.
Our definition is more general, but only places a cap on the formal quantum minimal
distance without ruling out global entanglement altogether.
%
%

\subsection{Overview of the proof}

Our main statement considers a $2$-complex
$\X = (X_0,X_1,X_2)$, with boundary operators $\partial_1,\partial_2$
such that $\X$ is $\eps$-pseudo-random, for some small $\eps > 0$.
We begin by defining a linear space $C \subseteq \F_2^n$ to be \emph{weakly-binomial} if its weight enumerator $(B_0,\dotsc,B_n)$ - i.e. the vector of $n+1$ bins, specifying
the number of words of $C$ in any given weight - this enumerator has an upper-bound that behaves like the binomial distribution:

\begin{definition}[Weakly-binomial subspace]\label{def:weak1}
A subspace $C\subseteq \F_2^n$ on $n$ bits is \emph{$(\zeta, \eta)$-weakly-binomial} if for some constants $\zeta>0$ and $\eta>0$ we have:
\begin{equation}
\forall k\in \{0,\hdots, n\}, \quad
B_k \leq \frac{2^{\zeta n}\binom{n}{k}}{|C^{\perp}|}+2^{\eta n},
\end{equation}
where $\{B_k\}$ is the weight enumerator of $C$ and $|C^{\perp}|$ is the size of the dual space of $C$.
\end{definition}

\noindent
Note that the normalization by $|C^{\perp}|$ is necessary, by setting $\zeta = \eta = 0$.
Our proof consists of two main steps:  
\begin{enumerate}
\item
We show that the subspace $C\subseteq \F_2^n$ spanned by the generators (hyper-edges) of an $\eps$-pseudorandom complex is weakly-binomial.
\item
We show that any weakly-binomial subspace $C$, for which the dual code $C^{\perp}$ also {\it contains} a LDPC code,
must have a relatively small dimension.
\end{enumerate}
Hence for a CSS code $(C_X, C_Z)$ corresponding to $\eps$-PR boundaries, the relative dimension of $C_X$ must be small, which implies the quantum code rate $k=n-\dim(C_X)-\dim(C_Z)$ must be large.
By standard distance-rate trade-offs in quantum error-correction this then implies an upper-bound on the
minimal quantum error-correcting distance of ${\cal C}$.

\subsubsection{An $\eps$-PR hyper-graph spans a weakly-binomial subspace}

In the first step, we would like to approximate the weight enumerator of a space $C_X$ 
spanned by a set of generators that satisfy
the pseudorandom condition
in \cref{def:ePR}.
The weight enumerator is approximated by considering a random walk ${\cal M}_1$ on the Cayley graph of the space $C_X$
using its set of LDPC-sparse, and $\eps$-pseudorandom set of generators.
The stationary distribution of ${\cal M}_1$, when summed-up over separate shells of $\F_2^n$ of fixed-weight
then provide exactly the weight enumerator.

We would hence like to ``project''-down ${\cal M}_1$ to a random walk ${\cal M}_2$ defined on $n+1$ nodes
corresponding to ``shells'' of fixed weight in $\F_2^n$.
Hence, we would like to define transition probabilities between ``weight-bins'' that are independent of which word we choose in a fixed-weight bin.  To do this, we define a coarse-graining of the chain over some fixed partition of the outcome space.  We choose the shells of fixed weight as the sets in our partition.  

So now, consider the line walk ${\cal M}_2$: it is comprised of $n+1$ bins, with non-zero transition probabilities
between nodes of distance at most $q$ - the locality of the generators.  We consider a bin $B_k$ - i.e. the set of words in $C_X$ of weight $k$, and ask:
suppose that we sample a uniformly random generator $g$ from the rows of $\partial_2$ and add it to $w\in B_k$:
what is the distribution of $|w + g|$?

Generically - this might be a hard problem to solve.
However, using the $\eps$-PR condition it becomes simpler:  this condition, when interpreted in the appropriate way,
tells us that the probability that $|w+g| = k+j$, where $j\leq q$, and $q$ is the locality of each generator - behaves
approximately like sampling a word of weight $q$ {\it uniformly at random} and adding it to $w$.
As a result, this implies that ${\cal M}_2$ assumes the form of a binomial chain, i.e. adding
uniformly random words of weight $q$, up to an {\it additive} error at most $\eps$.
We then analyze this chain, and show it implies that the stationary distribution of this perturbed
chain deviates from the pure unperturbed chain by a modest multiplicative exponential error, so long as the bins
we consider are ``close'' to the center $n/2$.
Far from the center bin $B_{n/2}$ we have no control - but this is translated to a small exponential additive error - 
which together implies weak binomiality.

\subsubsection{Weakly-binomial subspaces with LDPC duals have large dimension}

In this part of the proof we are given a subspace $C_X$, such that $C_X^{\perp}$ contains an LDPC code,
and $C_X$ is weakly binomial.
We think of $C_X$ as being the span of parity checks $C_X=S_x$ of a CSS code, and hence $C_X^{\perp}$ {\it contains}
also the space spanned by a set of LDPC parity checks $C_Z=S_z$.
We assume that the boundary operators are $d$-local, and each bit is incident on $K$ (Pauli $x$) checks,
for $d, K = O(1)$.

To place an upper-bound on the dimension of $C_X$ we invoke the Sloane Mac-Williams transform \cite{MS77}
which translates any weight enumerator on a code $C_X$, to the weight enumerator on the dual code $C_X^{\perp}$.
The {\it crux} of the argument is essentially a weak converse to previous results by Ashikhmin and Litsyn \cite{AL99}
which use the transform in the context of classical codes.

Consider for example a classical code of large distance.
It's weight enumerator $(B_0,\dotsc,B_n)$ is by definition such that $B_0 = 1$ (the zero word)
and $B_i = 0$ for all $0 < i \leq \delta_{\min}$.
The result by Ashikhmin and Litsyn shows that if this is the case, then the weight enumerator of the dual code
$(B_0^{\perp},\dotsc,B_n^{\perp})$ has an upper-bound that is very close to being binomial:
if one considers an interval of linear size around $n/2$, say $[n/3, \hdots{}, n/2,\hdots, 2n/3]$ then it is the case that
each $B_k$ is at most ${n\choose k} / |C_X|$ up to a multiplicative polynomial factor.

In our case, we consider a {\it quantum} code, namely a CSS code, and argue the opposite way:
we show that if the weight enumerator (of one of the corresponding classical codes) $(B_0,\dotsc,B_n)$ is {\it weakly} binomial
in the sense defined above, then the weight enumerator of the dual
$(B_0^{\perp},\dotsc,B_n^{\perp})$ does not have a precise prefix of $0$ bins, as in the classical case
of a large-distance code, but the first $\alpha n$ bins are still very small, for some constant $\alpha>0$.

On the other hand, and this happens only for the quantum case: we know that the dual code $C_X^{\perp}$
contains an LDPC code.
Any LDPC code - whether it is $\eps$-PR or not, has the property that in the appropriate scale,
the number of words in bin $B_k$ grows exponentially fast with $k$, at least for sufficiently small $k = \beta n$.

So together, we collide on the bins of the dual code the opposing forces of the upper-bound implied by the weak binomial distribution,
which implies that the lower-prefix of $B^{\perp}$ is very small, with the fact that this lower-prefix
blows-up exponentially fast because it contains an LDPC code.
This implies a stringent limit on the dimension of the parity check spaces - and hence a lower-bound on the
rate of the code, which in turn implies a stringent upper-bound on its minimal distance of the corresponding quantum code.


\section{Preliminaries}

\subsection{Notation}

We adopt the following conventions and notation throughout the paper. Even though some of the machinery we use (e.g.\ MacWilliams identity) hold in a more general setting, we restrict our attention to binary linear codes in the classical case and binary (qubit) codes based on the CSS construction in the quantum case. Consequently, all linear operators in this paper are over $\F_2$. We denote a binary quantum code that encodes $k$ qubits into $n$ qubits with distance $d_{\min}$ as $[[n, k, d_{\min}]]$. Similarly, a classical code that encodes $k$ bits into $n$ bits and has distance $d_{\min}$ will be denoted $[n, k, d_{\min}]$. We will write $\rho := k/n$ for the \emph{rate} of a code, classical or quantum, and $\delta_{\min} := d_{\min}/n$ for the \emph{error rate} of a code.


We will use $d$ to denote the largest size of a hyper-edge in a complex. Alternatively, for a code over $\F_2$ the parameter $d$ (not to be confused with the distance $d_{\min}$) represents the Hamming weight of its parity checks in the basis of minimal Hamming weight. The letter $K$ will be used to denote the degree of a vertex, i.e.\ it will denote the number of ($d-1$)-cells incident to each $d$ cell or, alternatively, the number of bits examined by each parity check. For $x \in \F_2^n$, $|x|$ denotes the Hamming weight of $x$. Finally, we define $\log(x) \equiv \log_2(x)$ and use $H(p) := -p \log p - (1-p) \log (1-p)$ to denote the \emph{binary entropy function}.

For a discrete set $E$ we use the notation $x\sim U[E]$ to denote that $x$ is a uniformly random element from $E$.

\begin{definition}
Let $g(n)$ and $h(n)$ be two functions of $n$. We write $g(n) \geq_p h(n)$ if, for all $n \geq 1$,
\begin{equation}
  g(n) \geq h(n) n^z
\end{equation}
for some constant $z$. Similarly, we write $g(n) \leq_p h(n)$ if, for all $n \geq 1$,
\begin{equation}
  g(n) \leq h(n) n^z.
\end{equation}
\end{definition}

\subsection{Kravchuk polynomials}

Kravchuk polynomials are a special set of orthogonal polynomials with many applications in error correction~\cite[p.~130]{MS77}.  They have a simple interpretation which makes their definition and many of their properties intuitive.


We fix $n$ to be some positive integer throughout.  Let $m \in \{0,\dotsc,n\}$ and denote by
\begin{equation}
  S_m := \{w \in \F_2^n : |w| = m\}
\end{equation}
the set of all length-$n$ strings of Hamming weight $m$.  Let $\chi_u: \F_2^n \to \{-1,+1\}$ be a character of $\F_2^n$ for some $u \in \F_2^n$, i.e.\ a function of the form $\chi_u(v) := (-1)^{u \cdot v}$ where $u \cdot v := \sum_{i=1}^n u_i v_i$ denotes the inner product modulo $2$.  The $m$-th Kravchuk polynomial evaluated at $x \in \{0, \dotsc, n\}$ is then defined as
\begin{equation}\label{eq:krav1}
  P_m(x)
  := \sum_{w \in S_m} \chi_u(w)
   = \sum_{w \in S_m} (-1)^{u \cdot w},
\end{equation}
where $u \in \F_2^n$ is any vector of Hamming weight $|u| = x$.  Note by symmetry that $P_m(x)$ does not depend on the word $u$ chosen as long as $|u|=x$.  Also note that $P_m(x)$ implicitly depends also on the dimension $n$ of the underlying space $\F_2^n$, which should be clear from the context.

For any integer $l \geq 0$ and formal variable $x$, we define the \emph{binomial coefficient} as the following degree-$l$ polynomial in $x$:
\begin{equation}
  \binom{x}{l} :=
    \frac{x(x-1) \dotsb (x-l+1)}{l!}. 
\end{equation}
For integers $l < 0$ the the binomial coefficient is taken to be zero.
Using this, Kravchuk polynomials can be written explicitly as follows:

\begin{definition}
The $m$-th \emph{Kravchuk polynomial}, for $m \in \{0, \dotsc, n\}$, is a degree-$m$ polynomial in $x \in \R$ given by
\begin{equation}\label{eq:krav2}
  P_m(x) := \sum_{l=0}^m (-1)^l \binom{x}{l} \binom{n-x}{m-l}.
\end{equation}
\end{definition}

\noindent
It is not hard to see that \cref{eq:krav1,eq:krav2} agree for integer values of $x$.

One of the most important properties of Kravchuk polynomials is that they are orthogonal under a particular inner product. This fact that can be easily verified using the above interpretation:

\begin{lemma}\label{lem:krav_orth}
For $i,j \in \{0,\dotsc,n\}$, the Kravchuk polynomials $P_i(k)$ and $P_j(k)$ satisfy
\begin{equation}
  \sum_{k=0}^n \binom{n}{k} P_i(k) P_j(k)= \delta_{ij} 2^n \binom{n}{i}
\end{equation}
where $\delta_{ij}$ is the Kronecker delta.
\end{lemma}


One important application of this orthogonality relation is that any polynomial $g(x)$ with $\deg(g) \leq n$ has a unique Kravchuk decomposition.  A simple way of determining such a decomposition is as follows:

\begin{fact}
If $g(x)$ is a polynomial of degree at most $n$, its Kravchuk decomposition is
\begin{equation}
  g(x) = \sum_{j=0}^n g_j P_j(x)
  \qquad
\end{equation}
where
\begin{equation}
  g_j := \frac{1}{2^n\binom{n}{j}}
         \sum_{k=0}^n \binom{n}{k} P_j(k) g(k).
\end{equation}
\end{fact}

Following the line of argument in~\cite{KL97}, we will make use of a particular decomposition for the polynomial $P_m(x)^2$:

\begin{lemma}
For any $m \in \{0,\dotsc,\floor{n/2}\}$,
\begin{equation}
  (P_m(x))^2 = \sum_{i=0}^m \binom{2i}{i} \binom{n-2i}{m-i} P_{2i}(x).
  \label{eq1}
\end{equation}
\end{lemma}

\begin{IEEEproof}
According to \cref{eq:krav1},
\begin{equation}
  (P_m(x))^2 = \sum_{w,w' \in S_m} (-1)^{u \cdot (w+w')}
  \label{eq:sq}
\end{equation}
for any $u \in \F_2^n$ such that $|u| = x$.  Note that $|w+w'| = 2i$ for some $i \in \{0,\dotsc,m\}$, so we can rewrite the right-hand side of \cref{eq:sq} as
\begin{equation}
  \sum_{i=0}^m \sum_{v \in S_{2i}} c_v (-1)^{u \cdot v},
\end{equation}
where the integer $c_v$ accounts for the number of ways two $n$-bit strings $w$ and $w'$ (each of Hamming weight $m$) can overlap to produce a given string $v = w + w'$ of weight $|v| = 2i$.  It is not hard to see that $c_v$ depends only on the Hamming weight of $v$ and is given by
\begin{equation}
  c_v = \binom{2i}{i} \binom{n-2i}{m-i}.
\end{equation}
Indeed, we simply need to account for all ways of splitting the $2i$ ones of $v$ into two groups of size $i$ each (one of the groups is contributed by $w$ while the other by $w'$) as well as picking $m-i$ out of the remaining $n-2i$ locations where the remaining $m-i$ ones of $w$ and $w'$ would cancel out.
\end{IEEEproof}

We will also need the following simple upper bound on Kravchuk polynomials:

\begin{lemma}\label{lemma:krav_upper}
For any $k \in \{0, \dotsc, n\}$,
\begin{equation}
  P_m(k) \leq \binom{n}{m}.
\end{equation}
\end{lemma}

\begin{IEEEproof}
This follows easily from \cref{eq:krav1}.  Let $u$ be any binary vector with $|u| = k$.  Then
\begin{align}
  P_m(k) = \sum_{w \in S_m} (-1)^{u \cdot w} \leq \sum_{w \in S_m} (1)^{u \cdot w} = \\
  \binom{n}{m}
\end{align}
as claimed.
\end{IEEEproof}

\subsection{The Sloane-MacWilliams transform}

We will use an important relation known as MacWilliams identity~\cite{MS77}.  Suppose we have some linear code $C$ in $\F_2^n$.  We define the weight enumerator of the code $C$ as a set of coefficients $\{B_k\}$ where each $B_k$ denotes the number of words of weight $k$ in the code.  The dual code is simply the code $C^\perp$, all the words in $\mathbb{F}_2^n$ that are orthogonal to all the words in $C$.  Of course the code $C^\perp$ has its own weight enumerator $\{B_k^\perp\}$.  We state these notions formally in the following definitions:

\begin{definition}\label{def:weight_enum}
Given a code $C$, and for all $k \in \{0, \dotsc, n\}$, we define the weight enumerator $B_k$ as:
\be 
B_k=\left|\{ x\in C : |x|=k  \} \right|
\ee
\end{definition}

We define the dual code:

\begin{definition}
Given a code $C$, we define the dual code $C^\perp$ as:
\be 
C^\perp=\{x \in \mathbb{F}_2^n : \forall c \in C , \, c\cdot{} x=0  \}
\ee
\end{definition}

Naturally the dual code has an analogously defined weight enumerator:

\begin{definition}\label{def:weight_enum_perp}
For a code $C$ and $C^\perp$, we define:
\be 
B_k^\perp=|\{ x\in C^\perp : |x|=k  \}
\ee
\end{definition}

The MacWilliams identity provides a way to write the weight enumerator of the dual code in terms of the weight enumerator of $\mathcal{C}$, and the Kravchuk polynomials. 
 
\begin{theorem}[\cite{MS77}]
 Let $\mathcal{C}$ be a linear code over $\F_2^n$ with weight enumerator $\{B_k\}$.  Denote the dual code $\mathcal{C}^\perp$ and its weight enumerator $\{B_k^\perp\}$.  Then, it holds that:
\be 
B_k^\perp= \frac{1}{|\mathcal{C}|}\sum_{j=0}^n P_{k}(j) B_j
\ee
where $|\mathcal{C}|$ denotes the number of codewords in $\mathcal{C}$.
\end{theorem}



\begin{lemma}[\cite{KL97}]\label{thm1}
Let $\{B_j\}$ be the weight enumerator of a code $\mathcal{C}$ on $n$ bits, and $\{B_j^\perp\}$ be the weight enumerator of its dual.  If $\alpha(x):=\sum_{j=0}^n \alpha_j P_j(x)$ for some coefficients $\alpha_j$, then
\begin{equation}
|C|\sum_{j=0}^n \alpha_j B_j^\perp=\sum_{j=0}^n \alpha(j) B_j.
\end{equation}
\end{lemma}

\begin{IEEEproof}
By the MacWilliams identity,
\begin{align}
\sum_{j=0}^n \alpha_j B_j^\perp=\sum_{j=0}^n \alpha_j \left[\frac{1}{|C|} \sum_{k=0}^n B_k P_j(k) \right]=\\
\nonumber \frac{1}{|C|}\sum_{k=0}^n B_k \sum_{j=0}^n\alpha_j P_j(k)=
\frac{1}{|C|}\sum_{j=0}^n B_j \alpha(j).
\end{align}
\end{IEEEproof}

For obvious reasons, this identity has many applications in error correction.  In particular it is a useful tool for establishing many bounds on quantum codes.  One such bound, which we will need in the proof of the main theorem, is provided by Ashikhmin and Litsyn:

\begin{proposition}[\protect{\cite[Corollary~2]{AL99}}]\label{prop:AL}
A quantum code with parameters $[[n, k, d_{\min}]]$ satisfies:
\begin{equation}
\frac{k}{n} \leq 1- \frac{\delta_{\min}}{2}\log(3)-H\left(\frac{\delta_{\min}}{2} \right) - o(1),
\end{equation}
\end{proposition}

\subsection{Markov chains}

We will use coarse-grained Markov chains in our analysis.  We wish to partition the discrete state space of a Markov chain and analyze the coarse-grained dynamics.  
Given a Markov chain $\mathcal{M}$ with state space $\Omega$, and some subset $A \subseteq \Omega$, for any probability distribution $\pi$ over $\Omega$ we will denote:
\be 
\pi_A:=\sum_{i \in A} \pi_i
\ee
We then define a coarse-graining of a Markov chain:

\begin{definition}[Coarse-grained Markov chain]\label{def:coarse_grain}
Let $\mathcal{M}$ be an irreducible Markov chain with state space $\Omega$.  Denote the probability of transitioning from $i$ to $j$ as $\mathcal{M}_{i, j}$.  Suppose we have a partition  $\{S_k\}$ of $\Omega$ (i.e.\ $\cup_k S_k=\Omega$ and $S_k \cap S_j=\emptyset$ for $k \neq j$).  Denote the Markov chain's stationary distribution by $\{\pi_j\}$.  
We denote the coarse-grained Markov chain with respect to $\{S_k\}$ and $\pi$ by $\mathcal{M}'$. It has exactly one state for each set in $\{S_i\}$.  If $A$ and $B$ are two sets in $\{S_i\}$ we define:
\be 
\mathcal{M}_{A, B}'=\sum_{i \in A} \sum_{j \in B} \frac{\pi_i}{\pi_A}\mathcal{M}_{i, j}.
\ee
\end{definition}

\begin{lemma}\label{lem:coarse2}
Let $\mathcal{M}$ be an irreducible Markov chain with stationary distribution $\{ \pi_j \}$, and suppose we have some partition of the state space $\{S_i\}$.  Suppose we construct the coarse-grained Markov chain $\mathcal{M}'$ with respect to the partition $\{ S_i \}$.  Denote the stationary distribution of $\mathcal{M}'$ as $\{\pi_{S_i}'\}$.  The stationary distribution of $\mathcal{M}'$satisfies:
\be 
\forall S_i \in \{S_i\} \,\,\,\,\pi_{S_i}'=\pi_{S_i}=\sum_{j \in S_i} \pi_j
\ee
\end{lemma}

\begin{IEEEproof}
Fix some $B \in \{S_i\}$, we can evaluate:
\be 
\sum_{S_i} \pi_{S_i} \mathcal{M}'_{S_i, B}=\sum_{S_i} \pi_{S_i} \sum_{j \in S_i} \sum_{k \in B} \frac{\pi_j}{\pi_{S_i}} \mathcal{M}_{j, k}
\ee
\begin{align} 
=\sum_{S_i} \sum_{j \in S_i} \sum_{k \in B} \pi_j \mathcal{M}_{j, k}=\sum_{j \in \Omega} \sum_{k \in B} \pi_j \mathcal{M}_{j, k}=\\
\nonumber\sum_{k \in B} \sum_{j \in \Omega} \pi_j \mathcal{M}_{j, k}
\end{align}
Since $\{\pi_j\}$ is stationary for the original chain, then
\be 
=\sum_{k \in B} \pi_k=\pi_B
\ee
So, the distribution $\{\pi_{S_i}\}$ is stationary for the coarse-grained chain ${\cal M}'$.
\end{IEEEproof}

\begin{definition}[Reversible Markov chains]
\noindent
Let ${\cal M}$ be a Markov chain on space $\Omega$ with stationary distribution $\pi$.
${\cal M}$ is said to be {\it reversible} if 
$$
\forall i,j\in \Omega, \quad
\pi_i {\cal M}_{i,j} = \pi_{j} {\cal M}_{j,i}
$$
\end{definition}

The following fact is standard: it says that the random walk on the Cayley graph of a finite group is reversible. This follows almost immediately from the fact that this walk is invariant under left multiplication by an element of the group: 

\begin{fact}\label{fact:Cayley}
Let $G$ be a group, and ${\cal G}(G,s)$ be the Cayley graph of $G$ w.r.t. some generating set $s\subseteq G$.
Let ${\cal M}$ denote the random walk on ${\cal G}$.
Then ${\cal M}$ is reversible.
\end{fact}

Next, we show that under our natural definition of coarse-graining, reversible chains remain reversible:

\begin{fact}\label{fact:reverse1}
Let ${\cal M}$ be a reversible Markov chain on space $\Omega$, with a stationary distribution $\pi$.
Let $\{S_k\}$ be some partition of $\Omega$.
Then the coarse-grained chain ${\cal M}'(\{S_k\},\pi)$ is reversible.
\end{fact}

\begin{IEEEproof}
Let $\mathcal{M}$ be a reversible Markov chain.  By definition, then,
\begin{equation}
\pi_i \mathcal{M}_{i, j}=\pi_j \mathcal{M}_{j, i}
\end{equation}
We can write:
\begin{align}
\pi_{S_i} \mathcal{M}_{S_i, S_j}=\pi_{S_i} \sum_{k_1 \in S_i} \sum_{k_2 \in S_j} \frac{\pi_{k_1}}{\pi_{S_i}}\mathcal{M}_{k_1, k_2}\\
\nonumber =\sum_{k_1 \in S_i} \sum_{k_2 \in S_j} \pi_{k_1} \mathcal{M}_{k_1, k_2}
\end{align}
by reversibility,
\begin{equation}
=\sum_{k_1 \in S_i}\sum_{k_2 \in S_j} \pi_{k_2} \mathcal{M}_{k_2, k_1}=\pi_{S_j}\mathcal{M}_{S_j, S_i}
\end{equation}
\end{IEEEproof}

\section{An $\eps$-PR complex spans a weakly-binomial subspace}

The goal of this section is to show that any degree-regular chain complex that is $\eps$-PR
yields a set of stabilizers that span a weakly-binomial space.
The main lemma of this section shows that $\eps$-pseudo-randomness (see \cref{def:ePR}) implies weak binomiality:

\begin{lemma}[$\eps$-PR hyper-edges span weakly-binomial space]\label{lem:pr2weak}
Let $G = (V,E)$ be a $d$-uniform hyper-graph, i.e. $E \subseteq {n\choose d}$ - subsets
of vertices of $V$ of size exactly $d$, where $deg(v) = K$ for each $v\in V$.
If $G$ is $\eps$-pseudorandom ($\eps$-PR), then the span of $E$ as words over $\F_2^n$
is $(\zeta, \eta)$ weakly binomial with constants $\zeta = \eps^{1/2}$ and $\eta = H(\eps^{1/(2d)})$.  
\end{lemma}


\subsection{Weight enumerators as stationary distributions}

We analyze the weight enumerator of $\eps$-PR complexes by associating them to natural Markov chains,
and then analyzing the stationary distributions of these Markov chains.

\begin{definition}
Let $G = (V,E)$ be a hyper-graph $|V| = n$, and let $\mathcal{C} \subseteq \F_2^n$ be the space spanned by the hyper edges of $G$ as vectors over $\F_2^n$.
We associate a pair of Markov chains to ${\cal C}$: ${\cal M}^1, {\cal M}^2$ as follows:
\begin{itemize}
\item
${\cal M}^1$ is the Markov chain defined by a random walk on to the Cayley graph of ${\cal C}$ using the set of generators $E$.
\item
Let $S_i$ be the set of words in $\mathcal{C}$ of weight $i$, formally defined as $S_i=\{x \in \mathcal{C} \,\, | \,\, |x|=i   \}$.  $\mathcal{M}^2$ is defined as the coarse-grained Markov chain with respect to the partition $\{S_i\}$,
and some stationary distribution of ${\cal M}^1$.
\end{itemize}
\end{definition}

Let us denote the Markov chains corresponding to the complete hyper-graph as $\mathcal{M}^1$ and $\mathcal{M}^2$ (in this case $E$ contains all words of weight $d$).  Denote the corresponding stationary distributions as $\pi^1$ and $\pi^2$. Similarly, let us denote the Markov chains corresponding to our $\eps$-pseudorandom complex as $\mathcal{M}^{1,\, \eps}$ and $\mathcal{M}^{2, \, \eps}$ and the corresponding stationary distributions as $\pi^{1,\, \eps }$ and $\pi^{2, \, \eps}$.  The main observation is that to understand the weight enumerator of $\mathcal{C}=span(E)$ for our $\eps$-pseudorandom complex, one can instead
look at the $1$-dimensional stationary distributions $\pi^{2, \,\eps}$ of ${\cal M}^{2, \, \eps}$.
It is easy to check that
\begin{equation}\label{eq:stat1}
\forall k\in [n], \quad
B_k = 2^m \pi^{2, \, \eps}_k.
\end{equation}

So now, if a $d$-uniform hyper-graph $G$ is $\eps$-pseudorandom we would like to characterize
its corresponding ${\cal M}^{1, \, \eps},{\cal M}^{2, \, \eps}$ Markov chains as an approximation of the Markov chains $\mathcal{M}^1$ and $\mathcal{M}^2$.  

\begin{proposition}\label{eq:epr1}
Let $G = (V,E)$, $|V| = n$, be a $d$-uniform hyper-graph which is $\eps$-PR.
Then we have in ${\cal M}^{2, \,\eps}$ the following transition probabilities for all $i\in [n], j\in [d]$:
\begin{align}\label{eq:mg}
\left|{\M^{2,\, \eps}_{i,i+d - 2j} - {d \choose j} (i/n)^{j} (1 - i/n)^{d-j}}\right| \leq \eps &.
\end{align}
\end{proposition}

\begin{IEEEproof}
Let $\pi^{1, \, \eps}$ denote the stationary distribution of ${\cal M}^{1, \, \eps}$.
Consider some $x \in S_i \subseteq \mathbb{F}_2^n$ and define $A \subseteq [n]$ as the subset of bits for which $x(i)=1$.  Let us choose $S$ in \cref{def:ePR} as $A$.  By the $\eps$-pseudorandom condition, the probability that $x + y\in S_{i+d-2j}$
where $y\sim U[E]$ is a value in the interval $\mathcal{T}(i, j)$ with 

\begin{equation}
\left|z-\binom{d}{j} \left( \frac{i}{n}\right)^{j} \left(1-\frac{i}{n} \right)^{d-j} \right|\leq \eps
\end{equation}
for all $z\in \mathcal{T}(i, j)$.

The Markov chain ${\cal M}^{2, \, \eps} $ is derived by a coarse-graining of ${\cal M}^{1, \, \eps}$ along shells $S_i$ with coefficients from $\pi^{1, \, \eps}$.
Hence, by definition of coarse-graining \cref{def:coarse_grain}
the transition probability in ${\cal M}^{2,\,\eps}$ between vertices $i$ and $i+d-2j$ is a convex combination of probability values in ${\cal T}(i,j)$.
Hence ${\cal M}^{2, \, \eps}_{i,i+d-2j}\in {\cal T}(i,j)$ as required.

%
%
\end{IEEEproof}

We note that when ${\cal M}^1$ is the Markov chain corresponding to the span of {\it all} generators of weight $d$ we have that the stationary distribution of 
its $1$-dimensional corresponding chain ${\cal M}^2$, $\pi^2$
satisfies
\begin{equation}\label{eq:stat2}
\forall k\in [n], \quad
\pi^2_k = 2^{-n} {n\choose k}.
\end{equation}

\subsection{Stationary distributions}

Having established that the transition probabilities of $\mathcal{M}^{2, \, \eps}$ are very close to the transition probabilities of $\mathcal{M}^2$, we prove that the stationary distribution $\pi^2$ is an upper bound on $\pi^{2, \, \eps}$, up to a modest exponential factor.

\begin{proposition}\label{prop:pr2weak_stationary}
\noindent
Let ${\cal M}^1$ be the Markov chain corresponding to the random walk on $\F_2^n$ using 
all words of weight $d$.
Let ${\cal M}^{1, \,\eps}$ denote the Markov chain
corresponding to the random walk on $\F_2^n$ using $E$.
Let ${\cal I}$ denote the interval $[n \eps^{1/(2d)}, n(1 -  \eps^{1/(2 d)})]$.
Let $\pi^1$ and $\pi^{1, \, \eps}$ denote the stationary distributions of $\M^1$ and $\M^{1, \, \eps}$,
respectively.
Let ${\cal M}^2, {\cal M}^{2, \, \eps}$ denote the coarse-graining of ${\cal M}^1, {\cal M}^{1, \, \eps}$,
to the $n+1$ shells $\{S_k\}_{k=0}^n$ using the stationary distributions $\pi^1, \pi^{1, \, \eps}$ respectively.
Let $\pi^2,\pi^{2, \, \eps}$ denote their corresponding stationary distributions.
Then
$$
\forall i\in {\cal I}, \quad
\pi^{2, \, \eps}_i \leq \pi^2_i \cdot 2^{2 \eps^{1/2} (n/2-i)}.
$$
\end{proposition}

\begin{IEEEproof}
Since both ${\cal M}^1, {\cal M}^{1, \, \eps}$ are random walks on finite Cayley graphs then by Fact \ref{fact:Cayley} they are reversible.
Hence by Fact \ref{fact:reverse1} the coarse-grained Markov chains ${\cal M}^1(\{S_k\},\pi^1), {\cal M}_1^{1, \, \eps}(\{S_k\},\pi^{1, \, \eps})$
are also reversible.
This implies
\be\label{eq:mpi1}
\forall i\in [n], \ 
{\cal M}^2_{i,i+1} \pi^2_i = 
{\cal M}^2_{i+1,i} \pi^2_{i+1}.
\ee
\be\label{eq:mpi2}
\forall i\in [n], \ 
{\cal M}^{2, \, \eps}_{i,i+1} \pi^{2, \, \eps}_i = 
{\cal M}^{2, \, \eps}_{i+1,i} \pi^{2, \, \eps}_{i+1}.
\ee
By definition of the interval ${\cal I}$ and by \cref{def:ePR}, any transition probability is lower bounded by $({\eps^{1/(2d)}})^d = \eps^{1/2}$ over $|j| \leq d$.
Hence:
\begin{equation}
\forall i\in {\cal I},|j|\leq d, \ \ 
\M_{i,i+j}^{2} \geq \eps^{1/2}.
\end{equation}
Therefore by definition \cref{def:ePR}
\be\label{eq:m1}
\forall i\in {\cal I},|j| \leq d \ 
\M_{i,i+j}^{2, \, \eps} \leq \M_{i,i+j}^2 \cdot (1 + \eps^{1/2}).
\ee
Together with Equations \ref{eq:mpi1},\ref{eq:mpi2} above this implies
\begin{align}\label{eq:61}
\forall i\in {\cal I}, \quad
\frac{\pi^{2, \, \eps}_i}{\pi^{2, \, \eps}_{i+1}}
\leq
\frac{\pi^2_i}{\pi^2_{i+1}}\cdot \frac{1 + \eps^{1/2}}{1 - \eps^{1/2}}
\leq\\
\nonumber \frac{\pi^2_i}{\pi^2_{i+1}}\cdot (1 + 2\eps^{1/2}),
\end{align}
where the last inequality follows from Taylor series expansion.
Thus,
\begin{align}
\frac{\pi^{2, \, \eps}_i}{\pi^{2, \, \eps}_{i+1}}
\cdot \ \ 
\hdots \ \ 
\cdot
\frac{\pi^{2, \, \eps}_{n/2-1}}{\pi^{2, \, \eps}_{n/2}}
\leq \\
\nonumber \frac{\pi^2_i}{\pi^2_{i+1}}
\cdot
\hdots
\cdot
\frac{\pi^2_{n/2-1}}{\pi^2_{n/2}}
\cdot
(1 + 2\eps^{1/2})^{(n/2 - i)}.
\end{align}
In addition, by monotonicity of the binomial distribution around $n/2$, we have that $\pi^2_{n/2} \geq n^{-k}$, for some constant $k>0$. 
Hence
\be
\forall i\in {\cal I}, \quad
\pi^{2, \, \eps}_i \leq 2^{2 \eps^{1/2} (n/2 - i)} n^k \pi^2_i.
\ee
\end{IEEEproof}

\subsection{Proof of \cref{lem:pr2weak}}

To describe briefly the proof of the lemma:  
Let $\mathcal{M}^1$ be the random walk using all words of weight $d$, and let $\mathcal{M}^{1, \, \epsilon}$ be the random walk using generators $E$. 
In general, the stationary distribution $\pi^{1, \, \eps}$ of the $n$-dimensional Markov chain ${\cal M}^{1, \, \eps}$ can very hard to analyze.
However, since we are only interested in the coarse graining of this distribution into $n+1$ fixed-weight Hamming shells $\pi^{2, \, \eps}$ - namely
the weight enumerator of $\mathcal{C}={\rm span}(E)$,
we can consider instead the coarse-grained Markov chain ${\cal M}^{2, \, \eps}$ - which is the coarse-graining of ${\cal M}^{1, \, \eps}$ using $\pi^{1, \, \eps}$.
This greatly simplifies the analysis in two senses:
On one hand - ${\cal M}^{2, \, \eps}$ is reversible - because it's the coarse-graining of the reversible chain ${\cal M}^{1, \, \eps}$.
On the other hand, by the $\eps$-PR condition, it is close to the $1$-dimensional chain defined by sampling uniformly at random a word of weight $d$, and adding it to a given word.
Together these two conditions imply a concise condition
on its stationary distribution $\pi^{2, \, \eps}$ - namely the weakly-binomial property.

\begin{IEEEproof}
Let $\mathcal{M}^1$ be the random walk using all words of weight $d$, and let $\mathcal{M}^{1, \, \epsilon}$ be the random walk using generators of the code.  Denote the stationary distributions similarly, as is described in the statement of \cref{prop:pr2weak_stationary}.  By \cref{eq:stat1} and \cref{eq:stat2}, the stationary distributions have the form:
\be 
\pi^{2, \, \eps}_j=\frac{B_j}{2^m}\,\,\,\,\,\,\,\,\,\,\,\,\, \pi^2_j=\frac{\binom{n}{j}}{2^n}
\ee
\cref{prop:pr2weak_stationary} 
implies that, for $j \in [n \eps^{1/(2d)}, n(1 -  \eps^{1/(2d)})]$:
\be 
\frac{B_j}{2^m} \leq \frac{2^{\eps^{1/2} n}\binom{n}{j}}{2^n}
\ee 
or
\be
B_j \leq \frac{2^{\eps^{1/2} n} \binom{n}{j}}{|C^\perp|}
\ee

For $j$ outside the error interval, we know nothing.  We can then write the general upper bound for all $j$ by including an additive ``error floor term''.  This error floor term has magnitude:
\be 
\binom{n}{n \eps^{1/(2d)}} \leq 2^{n H(\eps^{1/(2d)})}
\ee
So we can write the general upper bound as:
\be 
\forall j \in [0, \hdots{} n], \quad B_j \leq \frac{2^{\eps^{(1/2)} n}\binom{n}{j}}{|C^\perp|}+2^{n H(\eps^{1/(2d)})}.
\ee
Hence ${\cal C}$ is weakly-binomial with parameters 
$$
\zeta =  \eps^{1/2}, \quad \eta = H(\eps^{1/(2d)}).
$$
\end{IEEEproof}

\section{Weakly-binomial spaces have bounded minimal distance}

In this section we use the definition of weakly-binomial spaces to argue an upper-bound on the minimal distance of quantum codes.  The following lemma can be interpreted as demonstrating that QECCs with large distance are far from binomial.  

\begin{lemma}\label{lem:qecc2weak}
Let ${\cal Q} = (C_X, C_Z)_n$ be a family of quantum $[[n, k, d]]$ CSS codes as described in \cref{def:CSS},
where $\delta_{\min} = d/n>0$ is a constant independent of $n$.
Let $m_X=\dim(C_X)$ and $m_Z=\dim(C_Z)$, and suppose $m_X \geq m_Z$.
Suppose that $C_X,C_Z$ are spanned by generators of weight $d$, such that in the hypergraph of each of $C_X,C_Z$
the degree is some constant $K_1,K_2 = O(1)$.
If $C_X$ is $(\zeta, \eta)$ weakly-binomial for $\eta \leq 2^{-2 \log(2d)/d}$ and $\zeta < \eta^d$ then $\delta_{\min} \leq 4 H(\eta)$.
\end{lemma}

\noindent
\begin{IEEEproof}
Define
\begin{equation}\label{eq:f}
f_{d,\eta}(\gamma)=2 \gamma+(1-2 \gamma)H\left(\frac{\eta-\gamma}{1-2\gamma}\right)
\nonumber +\frac{1}{d^2}H\left(2\gamma d \right) - H(\eta)
\end{equation}
We use now \cref{fact:f} 
that analyzes $f_{d,\eta}$: by this fact, whenever $\eta \leq 2^{-2 \log(2d)/d}$ we have:
\be
h_{d,\eta} \equiv f_{d, \eta}(\eta^d) \geq \eta^d.
\ee
By assumption, $C_X$ is $(\zeta,\eta)$-weakly binomial, for $\zeta < \eta^d  \leq h_{d,\eta}$.
Fix $\gamma = \eta^d$, $t = \eta n, j = \gamma n$.
Consider the polynomial $P_t^2(x)$.
Using \cref{eq1} express $P_t^2(x)$ in the Kravchuk basis as:
\begin{equation}\label{eq:pt2}
P_t^2(k)=\sum_{i=0}^t \binom{2i }{i} \binom{n-2i}{t-i} P_{2i}(k),
\end{equation}
and denote
\begin{equation}\label{eq:alpha}
\alpha_{2i} :=  \binom{2i }{i} \binom{n-2i}{t-i}.
\end{equation}
Let $\{B_k\}$ be the weight enumerator of $C_X$ defined in \cref{def:weight_enum} and $\{B_k^\perp\}$ be the weight enumerator of $C_X^\perp$ defined in \cref{def:weight_enum_perp}.  
By Mac-Williams theorem \ref{thm1} and \cref{eq:pt2}:
\be
|C_X|\sum_{i=0}^t \alpha_{2i} B_{2i}^\perp 
=
\sum_{k=0}^n B_k P_t(k)^2 
\ee
Since the RHS is a weighted average of $B_k$ with positive coefficients, we can apply the upper bound we have on $B_k$ assuming weak-binomiality:
\begin{align} 
|C_X| \sum_{i=0}^t \alpha_{2i} B_{2i}^\perp \leq \sum_{k=0}^n \left[\frac{2^{\zeta n } \binom{n}{k}}{|C_X^\perp|}+2^{\eta n} \right]P_t(k)^2\\
\nonumber =\left[\sum_{k=0}^n \frac{2^{\zeta n}   \binom{n}{k} |C_X|}{2^n} P_t(k)^2 \right]+\left[\sum_{k=0}^n P_t(k)^2 2^{\eta n} \right]
\end{align}
Now apply \cref{eq:pt2} to the first term:
\begin{align}
=\left[\sum_{k=0}^n \frac{2^{\zeta n}   \binom{n}{k} |C_X|}{2^n} \sum_{i=0}^t \alpha_{2i}P_{2i}(k) \right]+
\nonumber \left[\sum_{k=0}^n P_t(k)^2 2^{\eta n} \right]
\end{align}
Reversing the order of summation yields:
\begin{align}
=\sum_{i=0}^t \frac{2^{\zeta n}   |C_X| \alpha_{2i} }{2^n}\sum_{k=0}^n \binom{n}{k}P_{2i}(k) + 
\nonumber \sum_{k=0}^n P_t(k)^2 2^{\eta n}
\end{align}
Now observe that we can interpret the inner sum as an inner product between $P_{2i}$ and $P_0$ - i.e. the constant function.  By \cref{lem:krav_orth}, $2^n \delta_{i,\, 0}$ must be zero unless $i=0$.
Thus, we have:
\be 
=\frac{2^{\zeta n}   |C_X| \alpha_{0} }{2^n}2^n + \sum_{k=0}^n P_t(k)^2 2^{\eta n}
\ee
Now we can apply \cref{lemma:krav_upper}
\be 
\leq 2^{\zeta n}   |C_X|\alpha_0+n\binom{n}{t}^2 2^{\eta n}
\ee
and so by Equation \ref{eq:alpha}
we derive the inequality:
\be 
|C_X| \sum_{i=0}^t \alpha_{2i}B_{2i}^\perp \leq 2^{\zeta n}   |C_X| \binom{n}{t}+n \binom{n}{t}^2 2^{\eta n}
\ee
Dividing by $|C_X|$,
\begin{align}
\sum_{i=0}^t \alpha_{2i}B_{2i}^\perp \leq 2^{\zeta n}    \binom{n}{t}+n \binom{n}{t}^2 \frac{2^{\eta n}}{|C_X|} \leq_p\\
 \nonumber 2^{\zeta n} \binom{n}{t}+\binom{n}{t}^2 \frac{2^{\eta n}}{|C_X|}
\end{align}
This implies for all $i \leq t$:
\be 
\alpha_{2i} B_{2i}^\perp \leq_p 2^{\zeta n} \binom{n}{t} +\binom{n}{t}^2 \frac{2^{\eta n}}{|C_X|}
\ee
and in particular since $t = \eta n$ then for $j=\gamma n=\eta^d \, n \leq t$ we have: 
\be 
\alpha_{2j} B_{2j}^\perp \leq_p 2^{\zeta n} \binom{n}{t} +\binom{n}{t}^2 \frac{2^{\eta n}}{|C_X|}
\ee
Applying \cref{eq:alpha}:
\be 
\binom{2 j}{j} \binom{n-2j}{t-j} B_{2j}^\perp \leq_p 2^{\zeta n} \binom{n}{t} +\binom{n}{t}^2 \frac{2^{\eta n}}{|C_X|}
\ee
By the definition of CSS codes (\cref{def:CSS}), $C_Z \subseteq C_X^\perp$ so we can lower bound the weight enumerator of $C_X^\perp$ with the weight enumerator of $C_Z$.  
Since by hypothesis $C_Z$ is also generated by a $d$-uniform hyper-graph of some fixed degree $K_2 = O(1)$, we can apply \cref{prop:low_bound}:
\begin{align} 
\binom{2j}{j}\binom{n-2 j}{t-j}\binom{\frac{n}{d^2}}{\frac{2j}{d}} \leq_p 
\nonumber 2^{\zeta n} \binom{n}{t}+\binom{n}{t}^2 \frac{2^{\eta n}}{|C_X|}
\end{align}
Next, by the binomial coefficient approximations in \cref{prop:bin}:
\begin{equation}\label{eq:52}
2^{2j+(n-2j)H\left(\frac{t-j}{n-2j}\right)+\frac{n}{d^2}H\left( \frac{2 j d}{n}\right)} \leq_p
\nonumber 2^{\zeta n+ n H\left(\frac{t}{n} \right)}+ 2^{n(2H(t/n)+\eta-m_X/n)}
\end{equation}
Now rewrite \cref{eq:52} in terms of $\gamma$ and $\eta$:
\begin{equation}\label{eq:45} 
2^{n\left(2 \gamma+(1-2 \gamma)H\left(\frac{\eta-\gamma}{1-2\gamma}\right)+\frac{1}{d^2}H\left(2\gamma d \right) \right)} \leq_p
\nonumber 2^{n\left(\zeta +H(\eta)\right)}+ 2^{n\left(2H(\eta)+\eta-m_X/n \right)}
\end{equation}
Now observe by that by our choice of parameters the first summand of RHS is negligible compared to LHS as follows:  
By assumption:
\be 
f_{d, \eta}(\eta^d)  = h_{d,\eta}  \geq \eta^d > \zeta.
\ee
or, by definition of $f_{d, \eta}$:
\begin{align}
2 \gamma+(1-2 \gamma)H\left(\frac{\eta-\gamma}{1-2\gamma}\right)+\frac{1}{d^2}H\left(2\gamma d \right) = 
\nonumber h_{d,\eta} +H(\eta) > \zeta + H(\eta).
\end{align}
Therefore, together with \cref{eq:45} this implies the following upper-bound:
\begin{align} 
\frac{1}{d^2}H(2 \gamma d) +2 \gamma + (1- 2\gamma) H\left( \frac{\eta-\gamma}{1-2\gamma} \right) \leq
\nonumber 2 H(\eta)+\eta - m_X/n
\end{align}
or 
\be 
f_{d, \eta}(\eta^d)-H(\eta) \leq \eta - m_X/n
\ee
or 
\be 
H(\eta)-f_{d, \eta} (\eta^d) \geq m_X/n - \eta
\ee
Since $f_{d, \eta} (\gamma) > \zeta > 0$, 
\begin{equation}\label{eq:84}
H(\eta) + \eta \geq m_X/n
\end{equation}
Since $m_X \geq m_Z$, 
\begin{equation}
1-\rho=\frac{m_X}{n}+\frac{m_Z}{n} \leq 2\frac{m_X}{n}
\end{equation}
or 
\be 
\frac{1-\rho}{2}\leq \frac{m_X}{n}
\ee
Hence, using equation \cref{eq:84}
\be 
\frac{1-\rho}{2} \leq H(\eta) + \eta
\ee
which implies
\be 
1-2 H(\eta) - 2 \eta \leq \rho
\ee
Now apply \cref{prop:AL}.  We obtain:
\begin{align}
1-2 H(\eta) - 2\eta \leq \\
\nonumber 1-\frac{\delta_{\min}}{2}\log(3) -H\left(\frac{\delta_{\min}}{2} \right)
\end{align}
It is easy to see that the RHS is a convex function of $\delta_{\min}$.  Clearly the linear function $\delta_{\min}$ is convex, and the binary entropy function is well known to be concave (hence $-H(x)$ is convex).  So, we can upper bound the RHS with a straight line  with endpoints at the beginning and end of our interval of interest.  Define:
\be 
g(\delta_{\min})=1-\frac{\delta_{\min}}{2}\log(3)-H\left( \frac{\delta_{\min}}{2} \right)
\ee
$g(0)=1$ and let $z>0$ be the first point such that $g(z)=0$.  It can be verified computationally that this fixed value of $z$  exists in the interval $[0.35, 0.4]$.  
So we can upper bound:
\be 
\forall \delta_{\min} \in [0, z] \,\,\,  g(\delta_{\min}) \leq 1-\frac{1}{z}\delta_{\min}
\ee
So, we can write:
\be 
1-2 H(\eta) - 2 \eta \leq 1-\frac{1}{z}\delta_{\min} 
\ee
Since $z < 1$, this implies: 
\be 
\delta_{\min} \leq 2 (H(\eta) + \eta).
\ee
By \cref{prop:entropy_low}, for $\eta \leq 1/2$ 
\be 
\delta_{\min} \leq 4 H(\eta)
\ee
\end{IEEEproof}

\begin{proposition}\label{prop:low_bound}
Let $C$ be an $m$-dimensional linear code over $\F_2$ on $n$ bits.  Suppose each generator of the code has weight $d$, and that the degree of every vertex is $K$ .  
For $k$ divisible by $d$ and $k\leq n/d$ we have:
\be 
B_k\geq \binom{\frac{n}{d^2}}{\frac{k}{d}}
\ee
\end{proposition}

\begin{IEEEproof}
The number of generators is $g = n K / d$.
We find a set of non-overlapping generators of maximal size.
By choosing a generator, and discarding all its neighbors we can find such a non-overlapping set of size at least $n/d^2$.
Consider any subset of $k/d$ such generators.
Any such set corresponds to a word in $C$ of weight exactly $k$.
Since $k\leq n/d$ then $k/d \leq n/d^2$ so the number of such words is therefore at least the binomial coefficient:
${n/(d^2) \choose k/d}$.
\end{IEEEproof}

\section{Proof of \cref{thm:main}}

\begin{IEEEproof}
Let $\X = (X_0,X_1,X_2)$ be a $2$-complex
and denote by ${\cal C}(\X) = [[n,k,d_{\min}]]$ be the associated quantum code, and write ${\cal C}(\X) = (C_X, C_Z)$.
Suppose that $\X$ is $\eps$-PR.
By Lemma \ref{lem:pr2weak} we have that $C_X$ is weakly-binomial with parameters $\zeta = \eps^{1/2}, \eta = H(\eps^{1/(2d)})$.
Since by assumption $\partial_1, \partial_2^T$ both have row weight $d$ we can invoke Lemma
\ref{lem:qecc2weak}.
The lemma states that if $\eta\leq 2^{-2\log(2d)/d}$ and $\zeta < \eta^d$ then $\delta_{\min} \leq 4H(\eta)$.
To check that this is indeed the case, observe that by assumption
$$
\eta  = H(\eps^{1/(2d)}) \leq 2^{-2 \log(2d)/d}.
$$
and if $\eps^{1/(2d)} \leq \frac{1}{2}$ or $\eps \leq 1/2^{2d}$.  So, by \cref{prop:entropy_low} we can write:
\begin{align}
\eta^d=\left[H(\epsilon^{1/(2d)})\right]^d > \\
\nonumber \eps^{(1/(2d))d}=\eps^{1/2}
\end{align}
The upper bound we obtain on the distance is:
\be 
\delta_{\min} \leq 4 H(H(\epsilon^{1/(2d)}))
\ee
For $\eta \leq 1/2$, applying \cref{prop:entropy_low}
\be 
\delta_{\min} \leq 4H(2 \epsilon^{1/(2d)}\log(1/\epsilon^{1/(2d)})) 
\ee
Applying it once more,
\begin{align}
\delta_{\min}=\frac{12}{d^2}\eps^{1/(2d)}\log^2(1/\eps)
\end{align}
\end{IEEEproof}

\subsection{Note on the proof}
The reader may have noticed that the ideal Markov chain $\mathcal{M}^1$ may not span all of $\mathbb{F}_2^n$, depending on the constant $d$. 

If $d$ is odd, then $\mathcal{M}^1$ must span all of $\mathbb{F}_2^n$, so our proof follows exactly as written.  This follows from a simple induction argument.  With any word of weight $d$, odd, we can make any word of weight $2$.  With any word of weight $2$ we can make any even weight word. Using any word of even weight and any word of weight $d$ allows us to make any word of any odd weight. 

Only minor modifications are needed to \cref{prop:pr2weak_stationary} if $d$ is even.  In this case the chain $\mathcal{M}^1$ only spans the even words, so we can only telescope on nonzero stationary probabilities in \cref{eq:61}.  Additionally, the inequality $\pi_1(n/2)\geq 1/n^k$ only holds up to a factor of $2$, not significantly effecting our bounds.
\section{Acknowledgments}

The authors wish to thank Tali Kaufman,  Alex Lubotzky, and Peter Shor for useful conversations.
LE was inspired in part by ideas presented at the ``Conference on High-Dimensional Expanders 2016'',
and thanks the organizers for inviting him.  LE is funded by NSF grant CCF-1629809, and the Templeton Foundation.
MO was funded by the Leverhulme Trust Early Career Fellowship (ECF-2015-256) and would like to thank MIT, where most of this work was done, for hospitality.  KT wishes to acknowledge the summer school at Weizmann Institute, ``High Dimensional Expanders, Inverse Theorems and PCPs'' for providing valuable insight into expanders.
KT is funded by the NSF through the STC
for Science of Information under grant number CCF0-939370.


\bibliographystyle{IEEEtran}
\bibliography{IEEEabrv,References}

\appendix

\section{Technical estimates}

\begin{fact}\label{fact:f}
Let
\begin{equation}
f_{d,\beta}(\gamma):=2 \gamma+(1-2 \gamma)H\left(\frac{\beta-\gamma}{1-2\gamma}\right)\\
\nonumber +\frac{1}{d^2}H\left(2\gamma d \right) - H(\beta)
\end{equation}
Then whenever $\beta \leq 2^{-2\log(2d)/d}$ we have:
$$
h_{d,\beta} \equiv f_{d,\beta}(\beta^d) \geq \beta^d.
$$
\end{fact}

\begin{IEEEproof}
Expanding $H\left(\frac{\beta-\beta^d}{1-2\beta^d}\right)$ to first order
$$
H\left(\frac{\beta-\beta^d}{1-2\beta^d}\right)
\geq
H(\beta) + \beta^d \cdot \log(\beta).
$$
Applying this inequality and discarding higher order terms in $\beta^d$ we obtain:
\begin{align}
f_{d,\beta}(\beta^d)\geq 
2 \beta^d 
+ \beta^d \log(\beta) 
+ 
\frac{1}{d^2}H\left(2\beta^d d \right) \\
\nonumber\geq
\beta^d \log(\beta) 
+ 
\frac{1}{d^2}H\left(2\beta^d d \right) \\
\nonumber\geq
\beta^d \log(\beta) - \frac{1}{d^2} (2\beta^d d) \log( 2\beta^d d)\\
\nonumber =
\beta^d \log(\beta) - \frac{1}{d^2} (2\beta^d d) (\log( \beta^d) + \log(2d)) \\
\nonumber =
- \beta^d \log(\beta) - \beta^d (2\log(2d)/d) \\
\nonumber \geq
- \frac{1}{2} \beta^d \log(\beta) \geq \beta^d
\end{align}
where the inequality before last assumes $\beta \leq 2^{-2\log(2d)/d}$.
\end{IEEEproof}

\noindent
The following fact appears in \cite[p.~427]{FJ11} and \cite[p.~121]{A90}.

\begin{proposition}\label{prop:bin}
Let $1 \leq k \leq n$.  If $\epsilon=k/n$, then
\be 
\frac{1}{\sqrt{8n \epsilon(1-\epsilon)}}2^{H(\epsilon)n} \leq \sum_{i=0}^k \binom{n}{i} \leq 2^{H(\epsilon)n}.
\ee
This implies that:
\be 
\binom{n}{k} \leq 2^{H(\epsilon)n}
\ee
and
\be 
\frac{1}{k\sqrt{8 n \epsilon(1-\epsilon)}}2^{n H(\epsilon)} \leq \binom{n}{k}
\ee
Or,
\be 
2^{n H(\epsilon)} \leq_p \binom{n}{\epsilon n} \leq_p 2^{n H(\epsilon)}
\ee
\end{proposition}

\begin{proposition}\label{prop:entropy_low}
The following holds for  all $x \in [0, 1/2]$,
\be 
x \leq H(x) \leq 2x\log(1/x)
\ee

\end{proposition}
\begin{IEEEproof}
The first inequality $x \leq H(x)$ holds since
\be 
H(x) \geq x\log(1/x) \geq x \,\, \forall x \in [0, 1/2]
\ee 
To prove the second inequality, define the function:
\be 
g(x)=x \log(1/x)-(1-x)\log(1/(1-x))
\ee
By taking the derivative and simplifying, its clear that the only local maximum of the function is at $x=1/2$.  Also, we can see that the derivative is positive at $x=1/4$, and that $g(0)=0=g(1/2)$.  The inequality follows.
\end{IEEEproof}

\end{document}